\documentclass[aps,prb,twocolumn,amsmath,amssymb,%
superscriptaddress,unsortedaddress,%
showpacs]{revtex4}

\usepackage{graphicx}
\usepackage{multirow}
\usepackage{dcolumn}% Align table columns on decimal point

\begin{document}

\title{Modified embedded-atom method interatomic potentials for the
  Mg-Al alloy system}

\author{B.~Jelinek}
\author{J.~Houze}
\author{Sungho~Kim}
\affiliation{
  Department of Physics and Astronomy, 
  Mississippi State University,
  MS State, Mississippi 39762, USA
}
\affiliation{%
Center for Advanced Vehicular Systems, 
Mississippi State University, 
Mississippi State, MS 39762, USA
}
\author{M.~F.~Horstemeyer}
\affiliation{
  Department of Mechanical Engineering,
  Mississippi State University,
  MS State, Mississippi 39762
}
\affiliation{%
Center for Advanced Vehicular Systems, 
Mississippi State University, 
Mississippi State, MS 39762, USA
}
\author{M.~I.~Baskes}
\affiliation{
  Los Alamos National Laboratory,
  MST--8, MS G755,
  Los Alamos, NM 87545
}
\author{Seong-Gon~Kim}\email{kimsg@hpc.msstate.edu} 
\affiliation{
  Department of Physics and Astronomy, 
  Mississippi State University,
  MS State, Mississippi 39762, USA
}
\affiliation{%
Center for Advanced Vehicular Systems, 
Mississippi State University, 
Mississippi State, MS 39762, USA
}

\date{\today}

\begin{abstract}

  We developed new modified embedded-atom method (MEAM) interatomic
  potentials for the Mg-Al alloy system using a first-principles
  method based on density functional theory (DFT). The materials
  parameters, such as the cohesive energy, equilibrium atomic volume,
  and bulk modulus, were used to determine the MEAM parameters.
  Face-centered cubic, hexagonal close packed, and cubic rock salt
  structures were used as the reference structures for Al, Mg, and
  MgAl, respectively.  The applicability of the new MEAM potentials to
  atomistic simulations for investigating Mg-Al alloys was
  demonstrated by performing simulations on Mg and Al atoms in a
  variety of geometries.  The new MEAM potentials were used to
  calculate the adsorption energies of Al and Mg atoms on Al (111) and
  Mg (0001) surfaces.  The formation energies and geometries of
  various point defects, such as vacancies, interstitial defects and
  substitutional defects, were also calculated. We found that the new
  MEAM potentials give a better overall agreement with DFT
  calculations and experiments when compared against the previously
  published MEAM potentials.

\end{abstract}

% PACS numbers
\pacs{%
61.50.Lt, %	Crystal binding; cohesive energy
62.20.Dc, %	Elasticity, elastic constants
61.72.Ji, % Point defects (vacancies, interstitials, color centers,
          % etc.) and defect clusters
68.35.-p 	% Solid surfaces and solid-solid interfaces: Structure and energetics
}

\maketitle

\section{Introduction}

Magnesium alloys are becoming increasingly important in many
technological areas, including aerospace and automotive industries.
The usage of magnesium die castings, for example, is increasing in the
automotive industry \cite{Han05:Mg-creep, Lou95:Mg-review,
  Pett96:Mg-review} due to the lower mass densities of magnesium
alloys compared with steel and aluminum, higher temperature
capabilities and improved crash-worthiness over plastics. The primary
magnesium alloys for die-casting applications are the
magnesium-aluminum alloys such as AM50 and AM60B \cite{Pekg92, Polm92,
  King98}.

To meet the industrial demand for high-strength light-weight magnesium
alloys, it is essential to obtain detailed understanding of the effect
of individual alloying elements on the properties of magnesium alloys,
especially among the main constituent elements, Mg and Al. The
alloying elements can form interstitial or substitutional defects, or
can precipitate into small particles creating complex interface
structures.  The interactions between these alloying elements need to
be investigated using atomistic simulation techniques such as
molecular dynamics or Monte Carlo simulations.  These atomistic
simulations require accurate atomic interaction potentials to compute
the total energy of the system.  First-principles calculations
certainly can provide the most reliable interatomic potentials.
However, realistic simulations of alloy systems often require a number
of atoms that renders these methods impractical -- they either require
too much computer memory or take too long to be completed in a
reasonable amount of time.  One alternative is to use (semi-)empirical
interaction potentials that can be evaluated efficiently, so that the
atomistic approaches that use them can, in certain cases, handle
systems with more than a million atoms.

There are two additional essential features that are expected from a
useful semiempirical approach besides its efficiency: reliability and
flexibility.  A reliable interatomic potential would accurately
reproduce various fundamental physical properties of the relevant
element or alloy, such as elastic, structural, and thermal properties.
Reliability also includes transferability.  A transferable interatomic
potential would perform reasonably well even under circumstances that
were not used during its construction phase.  A flexible semiempirical
approach can represent interaction potentials among a wide variety of
elements and their alloys using a common mathematical formalism.  The
modified embedded-atom method (MEAM) potential proposed by Baskes et
al. was the first semiempirical atomic potential using a single
formalism for fcc, bcc, hcp, diamond-structured materials and even
gaseous elements, in good agreement with experiments or
first-principles calculations \cite{baskes87:_applic,
  baskes89:_semiem, baskes92:_modif, baskes94:_meam_hcp}.  The MEAM is
an extension of the embedded-atom method (EAM) \cite{daw89:_model,
  daw84:_embed} to include angular forces.  The EAM was able to
reproduce physical properties of many metals and impurities. The EAM
was applied to hydrogen embrittlement in nickel
\cite{daw83:_semiem_hydrog_embrit}, and to nickel and palladium with
hydrogen \cite{daw84:_embed}. Cherne et al. made a careful
comparison of MEAM and EAM calculations in a liquid nickel
system\cite{Cherne:2001:PRB-65}.

Atomistic simulations of a wide range of elements and alloys have been
performed using the MEAM potentials.  \citet{baskes87:_applic} first
proposed the MEAM method to obtain realistic shear behavior for
silicon.  \citet{baskes89:_semiem} then provided the MEAM model of
silicon, germanium and their alloys.  The MEAM was also applied to 26
single elements\cite{baskes92:_modif} and to silicon-nickel alloys and
interfaces\cite{baskes94:_atomis}.  \citet{gall00:_atomis} used the
MEAM to model the tensile debonding of an aluminum-silicon interface.
\citet{lee00:_secon} improved the MEAM to account for the second
nearest-neighbor interactions. Also, \citet{huang95:_molec} used the
MEAM and two other potentials to determine defect energetics in
beta-SiC.  The MEAM parameters for a nickel and molybdenum-silicon
system were determined by \citet{baskes97:_deter,baskes99:_atomis}.
Recently, an effort has been made by \citet{lee03:_semiem_cu_ag_au_ni}
to create the MEAM potentials for Cu, Ag, Au, Ni, Pd, Pt, Al, and Pb,
based on the first and the second nearest-neighbor MEAM.  A new
analytic modified embedded-atom method (AMEAM) many-body potential was
also proposed and applied to 17 hcp metals, including
Mg\cite{hu01:_analy_hcp,hu03:_point_hcp}.  For the Mg-Al alloy system,
a set of EAM potentials have been developed using the ``force
matching'' method by \citet{Liu:SURF-v373-1997}.  The structural
properties of various polytypes of carbon were described using a MEAM
potential\cite{lee05:_carbon}.  Finally, \citet{potir06:_molec} used
the MEAM to analyze damage evolution in a single crystal nickel.

The purpose of the present work is to develop the MEAM potentials for
aluminum, magnesium, and their alloy systems based on first-principles
calculations using density-functional theory (DFT). Energy
calculations and geometry optimizations of various structures were
performed within the local-density approximation (LDA)
\cite{Ceperley:1980, Perdew:1981} using ultrasoft pseudopotentials
\cite{Vanderbilt:1990, Kresse:1994, Kresse:1996:VASP:PRB-69}.  The
cross pair potential was constructed by fitting elastic properties
from DFT calculations for aluminum and magnesium in the B1 reference
structure.  First, the equilibrium lattice parameter, cohesive energy,
bulk modulus, trigonal and tetragonal shear moduli were determined
from DFT calculations.  The pair potential was then constructed to fit
the equilibrium volume and bulk modulus from \textit{ab initio}
calculations.  Moreover, an effort has been made to match the sign of
trigonal and tetragonal shear moduli.  The new MEAM potentials were
used to find the most energetically favorable structures for single
elements and their pair combinations.  The resulting energy-volume
curves reasonably match the \textit{ab initio} calculations.
Satisfactory agreement of vacancy formation and stacking fault
energies from DFT and MEAM calculations was found.  Throughout this
paper, the performance of our new potentials will be compared with the
previously published MEAM potentials\cite{lee03:_semiem_cu_ag_au_ni,
  hu01:_analy_hcp, Liu:SURF-v373-1997}.

The paper is organized in the following manner. In
Sec.~\ref{sec:Theory}, we give a brief review of the MEAM. In
Sec.~\ref{sec:Procedure}, the procedure for determination of the MEAM
parameters is presented along with the new MEAM interatomic potential
parameters. Validation of the newly developed MEAM potentials is
presented in Sec.~\ref{sec:Validation}. Different bulk structures,
surface defects, and point defects calculations were performed and
compared with DFT calculations and experiments.  Finally, in
Sec.~\ref{sec:Conclusion}, we discuss and summarize the results.

\section{MEAM theory}
\label{sec:Theory}

The total energy $E$ of a system of atoms in the
MEAM\cite{kim06:_meam_ti_zr} is approximated as the sum of the atomic
energies
\begin{equation}
  E = \sum_{i} E_i.
\end{equation}
The energy of atom $i$ consists of the embedding energy and the pair
potential terms:
\begin{equation}
  E_i = F_i\left( \bar\rho_{i} \right) + 
  \frac{1}{2} \sum_{j \neq i}\phi_{ij}\left(r_{ij}\right).
\end{equation}
$F$ is the embedding function, $\bar\rho_{i}$ is the background
electron density at the site of atom $i$, and
$\phi_{ij}\left(r_{ij}\right)$ is the pair potential between atoms $i$
and $j$ separated by a distance $r_{ij}$.  The embedding energy
$F_i\left(\bar\rho_{i}\right)$ represents the energy cost to insert
atom $i$ at a site where the background electron density is
$\bar\rho_{i}$. The embedding energy is given in the form
\begin{equation}
  \label{eq:emb}
  F_i\left(\bar\rho_{i}\right) = 
  A_{i} E_{i}^{0} \bar\rho_{i} \ln \left(\bar\rho_i\right),
\end{equation}
where the sublimation energy $E_i^0$ and parameter $A_i$ depend on the
element type of atom $i$.  The background electron density
$\bar\rho_i$ is given by
\begin{equation}
  \bar\rho_{i} = \frac{\rho_{i}^{\left( 0 \right)}}{\rho_{i}^0}
  G\left( \Gamma_i \right),
\end{equation}
where
\begin{equation}
  \Gamma_i = \sum_{k=1}^3 t_i^{\left(k\right)}
  \left(
    \frac{\rho_i^{\left(k\right)}}{\rho_i^{\left(0\right)}}
  \right)^2
\end{equation}
and
\begin{equation}
  G(\Gamma) = \sqrt{1 + \Gamma}.
\end{equation}
The zeroth and higher order densities, $\rho_i^{(0)}$, $\rho_i^{(1)}$,
$\rho_i^{(2)}$, and $\rho_i^{(3)}$ are given in
Eqs.~(\ref{eq:part_den}).  The composition-dependent electron density
scaling $\rho_i^0$ is given by
\begin{equation}
  \rho_i^0 = \rho_{i0}Z_{i0}G\left( \Gamma_i^\text{ref} \right),
\end{equation}
where $\rho_{i0}$ is an element-dependent density scaling, $Z_{i0}$
is the first nearest-neighbor coordination of the reference system, and
$\Gamma_i^\text{ref}$ is given by
\begin{equation}
  \Gamma_i^\text{ref} = \frac{1}{Z_{i0}^2}
  \sum_{k=1}^3 t_i^{\left( k \right)} s_i^{\left( k \right)},
\end{equation}
where $s_i^{(k)}$ is the shape factor that depends on the reference
structure for atom $i$. Shape factors for various structures are
specified in the work of \citet{baskes92:_modif}.  The partial
electron densities are given by
\begin{subequations}
  \label{eq:part_den}
\begin{eqnarray}
  \label{eq:part_den_first}
  \rho_i^{\left( 0 \right)} & = &
  \sum_{j \neq i} \rho_j^{a\left( 0 \right)} \left( r_{ij} \right) S_{ij}\\
  \left( \rho_i^{\left( 1 \right)} \right)^2 & = &
  \sum_{\alpha}
  \left[
    \sum_{j \neq i} \rho_j^{a\left( 1 \right)}
    \frac{r_{ij\alpha}}{r_{ij}} S_{ij}
  \right]^2\\
  \left( \rho_i^{\left( 2 \right)} \right)^2 & = &
  \sum_{\alpha, \beta}
  \left[
    \sum_{j \neq i} \rho_j^{a\left( 2 \right)}
    \frac{r_{ij\alpha}r_{ij\beta}}{r_{ij}^2} S_{ij}
  \right]^2\nonumber\\
  & - & \frac{1}{3}
  \left[
    \sum_{j \neq i} \rho_j^{a\left( 2 \right)}
    \left( r_{ij} \right) S_{ij}
  \right]^2
  \\
  \left( \rho_i^{\left( 3 \right)} \right)^2 & = &
  \sum_{\alpha, \beta, \gamma}
  \left[
    \sum_{j \neq i} \rho_j^{a\left( 3 \right)}
    \frac{r_{ij\alpha}r_{ij\beta}r_{ij\gamma}}{r_{ij}^3} S_{ij}
  \right]^2\nonumber\\
  & - & \frac{3}{5} \sum_{\alpha}
  \left[
    \sum_{j \neq i} \rho_j^{a\left( 3 \right)}
    \frac{r_{ij\alpha}}{r_{ij}} S_{ij}
  \right]^2,
  \label{eq:part_den_last}
\end{eqnarray}
\end{subequations}
where $r_{ij\alpha}$ is the $\alpha$ component of the displacement vector
from atom $i$ to atom $j$.  $S_{ij}$ is the screening function between
atoms $i$ and $j$ and is defined in Eqs.~(\ref{eq:scr}).  The atomic
electron densities are computed as
\begin{equation}
  \rho_i^{a\left( k \right)}\left( r_{ij}\right) =
  \rho_{i0} \exp
  \left[
    - \beta_i^{\left( k \right)} \left( \frac{r_{ij}}{r_i^0} - 1 \right)
  \right],
\end{equation}
where $r_i^0$ is the nearest-neighbor distance in the single-element
reference structure and $\beta_i^{\left( k \right)}$ is
element-dependent parameter.
Finally, the average weighting factors are given by
\begin{equation}
  t_i^{\left( k \right)} = 
  \frac{1}{\rho_i^{\left( 0 \right)}}
  \sum_{j \neq i} t_{0, j}^{\left( k \right)} \rho_j^{a\left( 0 \right)} S_{ij},
\end{equation}
where $t_{0,j}^{\left( k \right)}$ is an element-dependent parameter.

The pair potential is given by
\begin{align}
  \label{eq:pair}
  \phi_{ij}\left(r_{ij}\right) &= \bar\phi_{ij}\left(r_{ij}\right)
  S_{ij}\\
  \begin{split}
  \bar\phi_{ij}\left(r_{ij}\right) &= \frac{1}{Z_{ij}}
  \left[ 2E_{ij}^u \left( r_{ij} \right) 
    -F_i\left(\frac{Z_{ij}}{Z_i} \rho_j^{a(0)} 
  \left( r_{ij} \right) \right) \right. \\
  &\quad \left. -F_j\left(\frac{Z_{ij}}{Z_j} \rho_j^{a(0)} 
  \left( r_{ij} \right) \right) \right] \label{eq:rhohat}
\end{split} \\
  E_{ij}^u\left( r_{ij} \right) 
  &= -E_{ij}\left( 1 + a_{ij}^*\left(r_{ij}\right)\right)
  e^{-a_{ij}^{*}\left(r_{ij}\right)}\\
  a_{ij}^{*} &= \alpha_{ij} \left( \frac{r_{ij}}{r_{ij}^0} - 1 \right),
\end{align}
where $E_{ij}$, $\alpha_{ij}$ and $r_{ij}^0$ are element-dependent
parameters and $Z_{ij}$ depends upon the structure of the reference
system. The background densities $\hat\rho_i(r_{ij})$ in
Eq.~(\ref{eq:rhohat}) are the densities for the reference structure
computed with interatomic spacing $r_{ij}$.

The screening function $S_{ij}$ is designed so that $S_{ij} = 1$ if
atoms $i$ and $j$ are unscreened and within the cutoff radius $r_c$,
and $S_{ij} = 0$ if they are completely screened or outside the cutoff
radius. It varies smoothly between 0 and 1 for partial screening. The
total screening function is the product of a radial cutoff function
and three body terms involving all other atoms in the system:
\begin{subequations}
  \label{eq:scr}
  \begin{align}
    \label{eq:scr_first}
    S_{ij} &= \bar S_{ij} f_c \left( \frac{r_c - r_{ij}}{\Delta r} \right)\\
    \bar S_{ij} &= \prod_{k\ne i,j}S_{ikj}\\
    S_{ikj} &= f_c \left(\frac{C_{ikj} - C_{\text{min},ikj}}
      {C_{\text{max},ikj} - C_{\text{min},ikj}} \right)\\
    C_{ikj} &= 1 + 2 \frac{r_{ij}^2 r_{ik}^2 + r_{ij}^2 r_{jk}^2 
      - r_{ij}^4}{r_{ij}^4 - \left( r_{ik}^2 - r_{jk}^2 \right)^2}\\
    f_c\left(x\right) &=
    \begin{cases}
      1 & x \geq 1\\
      \left[ 1 - \left( 1 - x )^4 \right) \right]^2 & 0<x<1\\
      0 & x \leq 0\\
    \end{cases}
    \label{eq:scr_last}
  \end{align}
\end{subequations}
Note that $C_{\text{min}}$ and $C_{\text{max}}$ can be defined
separately for each $i$-$j$-$k$ triplet, based on their element
types. The parameter $\Delta r$ controls the distance over which the
radial cutoff is smoothed from 1 to 0 near $r=r_c$.

\section{Determination of the MEAM potential parameters}
\label{sec:Procedure}

\subsection{MEAM potentials for pure Al and Mg system}

The previously published MEAM parameters for
Al\cite{lee03:_semiem_cu_ag_au_ni} and Mg\cite{baskes92:_modif} served
as the basis for the present work. For some of the surface and point
defect calculations, however, we observed less than satisfactory
agreement between DFT and the MEAM calculations when these original
MEAM potential parameters were used. We followed similar procedures
prescribed by \citet{lee03:_semiem_cu_ag_au_ni} and
\citet{baskes92:_modif} to fine-tune the parameters and improve the
overall agreement with experiments and DFT calculations.  For each
element, several material parameters obtained from the reference
structure are utilized to determine the model parameters.  These
materials parameters include the cohesive energy, equilibrium atomic
volume, bulk modulus, and several elastic constants.  The most stable
crystal structures were chosen as the reference structures, namely a
face-centered cubic (fcc) structure for Al and a hexagonal close
packed (hcp) structure for Mg.  The new parameters obtained from the
present work are listed in Table~\ref{tab:meam_Mg_Al}.  The merits of
these new potentials are demonstrated in various calculations
described in Sec.~\ref{sec:Validation}.
\begin{table*}[!tbp]
  \caption{\label{tab:meam_Mg_Al} Set of the MEAM potential parameters 
    for pure Al and Mg. 
    $E_\text{c}$ is the cohesive energy, $a_0$ is the equilibrium lattice
    parameter, $A$ is the scaling factor for the embedding energy,
    $\alpha$ is the exponential decay factor for the universal energy,
    $\beta^{(0-3)}$ are the exponential decay factors for the atomic
    densities, $t^{(0-3)}$ are the weighting factors for the atomic
    densities, $C_{\text{max}}$ and $C_{\text{min}}$ are the screening
    parameters.  The reference structures for Al and Mg are fcc and
    hcp, respectively.}
  \begin{ruledtabular}
    \begin{tabular}{ccccccccccccccc}
      Element &
      $E_\text{c}$[eV] & $a_0$[\AA] & $A$ & $\alpha$ &
      $\beta^{(0)}$ & $\beta^{(1)}$ & $\beta^{(2)}$ & $\beta^{(3)}$ &
      $t^{(0)}$ & $t^{(1)}$ & $t^{(2)}$ & $t^{(3)}$ &
      $C_{\text{max}}$ & $C_{\text{min}}$\\
      \hline
      Al    &
      3.353 & 4.05 &  1.07 & 4.64 &
      2.04  & 1.50 &  6.0  & 1.50 &
      1.00  & 4.00 & -2.30 & 8.01  &
      2.8   & 2.0  \\
      Mg    &
      1.55  & 3.20 &  1.11 & 5.45 &
      2.70  & 0.0  &  0.35 & 3.0  &
      1.00  & 8.00  & 4.10 &-2.00  &
      2.8   & 2.0  \\
    \end{tabular}
  \end{ruledtabular}
\end{table*}

\subsection{MEAM potential for the Mg-Al alloy system}

The parameters of the MEAM potential for the Mg-Al alloy system were
determined from a procedure similar to the one prescribed by
\citet{Lee05:Fe-C}.  The parameters were constructed to fit the
elastic properties obtained from the DFT calculations for MgAl in the
rock-salt (B1) structure, which was chosen to be the reference
structure.  The new parameters obtained from the present work are
listed in Table~\ref{tab:meam_Mg-Al_alloy}.  Primary emphasis was put
on matching the equilibrium volume and the bulk modulus, which were
reproduced exactly (see Table~\ref{tab:Mg-Al-in-B1}).
\begin{table}[!tbp]
  \caption{\label{tab:meam_Mg-Al_alloy} The MEAM
    potential parameters for the Mg-Al alloy system.
    $E_c$ is the cohesive energy, $r_e$ is the equilibrium nearest
    neighbor distance, $\alpha$ is exponential decay factor for the
    universal energy, $C_{\text{max}}$ and $C_{\text{min}}$ are
    screening parameters, $\rho_{0}$ is the density scaling factor.
  }
  \begin{ruledtabular}
    \begin{tabular}{ccc}
      Parameter & Value\\
      \hline
      $E_c$[eV]  & 
      $(E_c^\text{Al}+E_c^\text{Mg})/2 - 0.4575$ \\
      $r_e$[\AA] & 2.821 \\
      $\alpha$         & 4.915  \\
      $C_{\text{min}}$(Al--Mg--Al) & 0.0 \\
      $C_{\text{min}}$(Mg--Al--Mg) & 2.0 \\
      $C_{\text{min}}$(Al--Al--Mg) & 2.0 \\
      $C_{\text{min}}$(Al--Mg--Mg) & 2.0 \\
      $C_{\text{max}}$(Al--Mg--Al) & 2.8 \\
      $C_{\text{max}}$(Mg--Al--Mg) & 2.8 \\
      $C_{\text{max}}$(Al--Al--Mg) & 2.8 \\
      $C_{\text{max}}$(Al--Mg--Mg) & 2.8 \\
      $\rho_{0} (\text{Al})$  & 1.0   \\
      $\rho_{0} (\text{Mg})$  & 0.6   \\
    \end{tabular}
  \end{ruledtabular}
\end{table}

\section{Validation of MEAM potentials}
\label{sec:Validation}

We demonstrate the validity and the transferability of the new MEAM
potentials by performing simulations on Al and Mg atoms in a variety
of structural arrangements.

\subsection{Bulk}

\subsubsection{Pure Al and Mg system}

\begin{figure}[!tbp]
  \includegraphics[width=1\columnwidth]{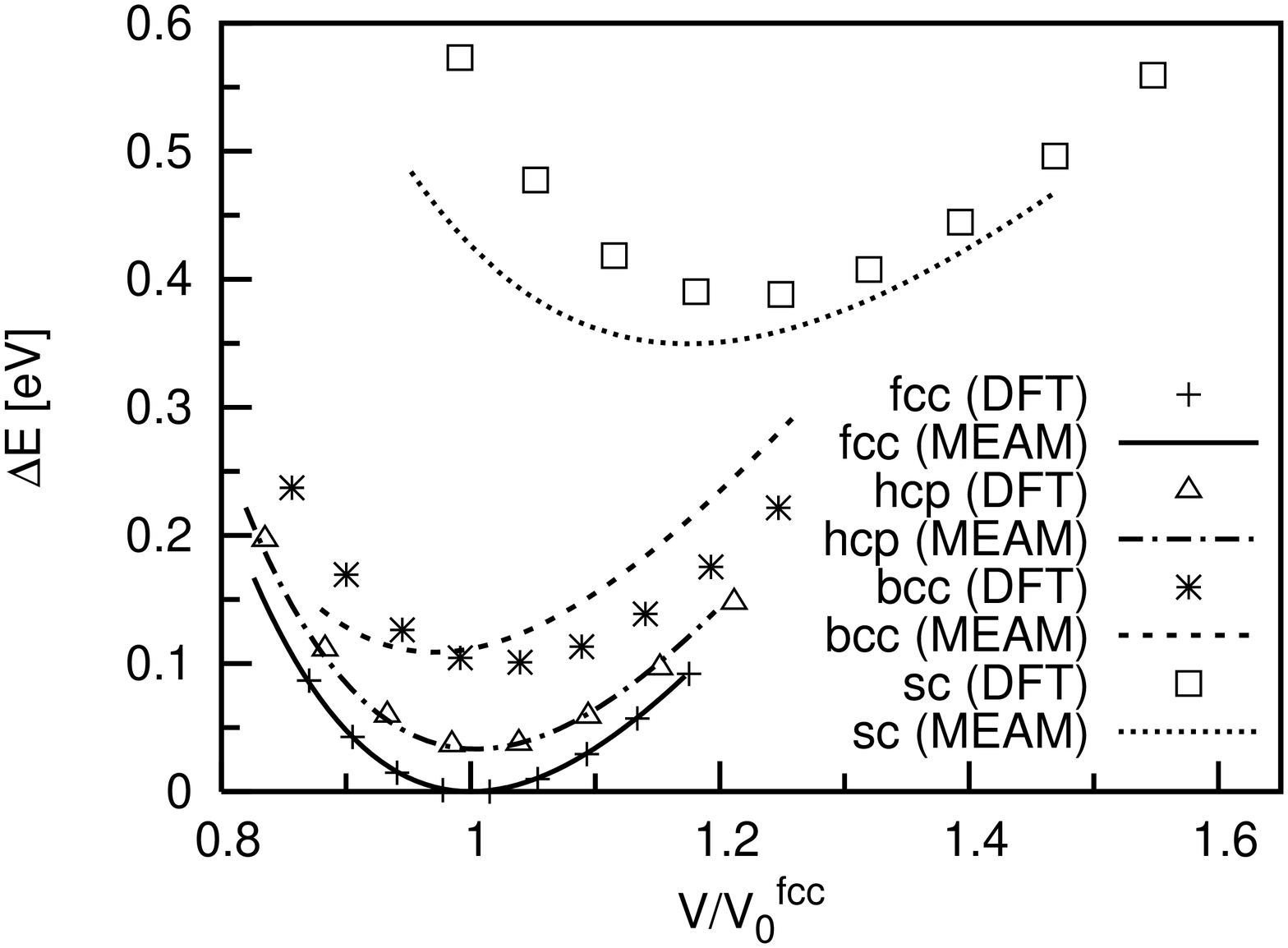}
  \caption{\label{fig:Al_struct} Atomic energies (total energies per
    atom) as a function of the atomic volume (volume per atom) for Al
    atoms in fcc, hcp, bcc and simple cubic (sc) crystal structures.
    The energies are measured from the equilibrium atomic energy of
    fcc structure.  Volumes are scaled by the equilibrium atomic
    volume of the fcc structure $V_0^{\text{fcc}}$.}
\end{figure}

To test the validity of the MEAM potentials for single elements, each
element was put into fcc, hcp, body-centered cubic (bcc), and simple
cubic (sc) crystal structures. The atomic energies for several atomic
volumes near equilibrium atomic volume were calculated.  The results
were compared with those of DFT calculations, as shown in
Fig.~\ref{fig:Al_struct}.
As expected, the curve for the fcc structure produced by the MEAM
potential retraces the results of DFT calculations nearly perfectly
since fcc was used as the reference structure during the potential
construction process. The agreement between the MEAM potential and DFT
for the hcp structure is also remarkable. The most important result,
however, is the fact that the new MEAM potential correctly identified
fcc as the most stable structure for Al.  Furthermore, the sequence of
the structures is correctly predicted in the order of stability by the
new Al MEAM potential. The relative cohesive energies, with respect to
the one for the fcc structure, are also in good agreement with the DFT
calculations, although the result for the simple cubic structure is
slightly underestimated. The relative equilibrium atomic volumes, with
respect to the one for the fcc structure, are also well reproduced. We
point out that the equilibrium atomic volume for fcc Al, obtained by
the MEAM potential (16.61\AA$^3$), is slightly different from the one
predicted by DFT (15.76\AA$^3$). This is due to the fact that the MEAM
parameters are fitted to reproduce the experimental volume, while DFT
within LDA tends to underestimate the equilibrium lattice constants by
roughly 1\% (see, e.g.  Ref.~\onlinecite{Vanderbilt:1994}).

\begin{figure}[!tbp]
  \includegraphics[width=1\columnwidth]{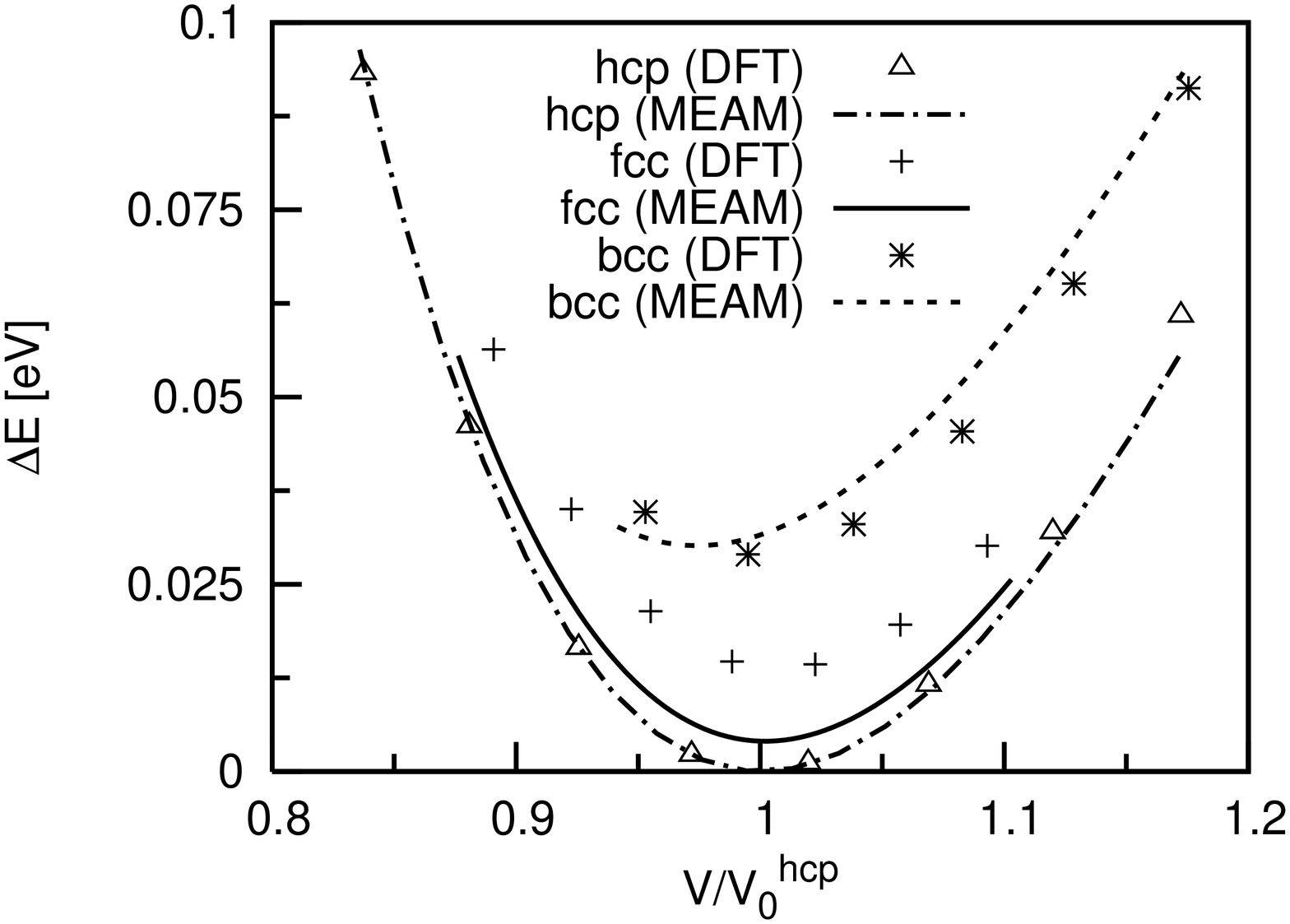}
  \caption{\label{fig:Mg_struct} Atomic energies of Mg as a function
    of the atomic volume in fcc, hcp, and bcc cubic crystal
    structures. The energies are measured from the equilibrium atomic
    energy of the hcp structure.  Volumes are scaled by the
    equilibrium atomic volume of the hcp structure $V_0^{\text{hcp}}$.
  }
\end{figure}
Fig.~\ref{fig:Mg_struct} shows the atomic energy plot for Mg atoms in
different crystal structures compared with the results of the DFT
calculations.  The hcp structure was used as the reference structure
for the Mg MEAM potential, and the DFT data points for this structure
are accurately reproduced.  The sequence of the structures is again
predicted correctly in the order of stability by the new Mg MEAM
potential. The relative atomic energies, with respect to the one for
the hcp structure, are also in good agreement with the DFT
calculations. Note that the scale of the vertical axis of
Fig.~\ref{fig:Mg_struct} is six times larger than that of
Fig.~\ref{fig:Al_struct}, and that the largest error in relative
atomic energies (fcc case) is in the order of 0.01 eV.  Similar to the
Al potential, equilibrium atomic volume for hcp Mg in the MEAM is set
to the experimental value of 23.16\AA$^3$, while DFT predicts a
smaller value of 21.54\AA$^3$.  Both the MEAM and the DFT methods
prefer a $c/a$ ratio close to 0.994 of the ideal $c/a$ ratio.

\subsubsection{Mg-Al alloy system}

To compare Mg-Al alloy systems with different stoichiometric
coefficients, we define the heat of formation per atom as
\begin{equation}
  H_{\text{f}} = \frac{E_{\text{tot}} - N_{\text{Mg}}\varepsilon_{\text{Mg}} 
    - N_{\text{Al}}\varepsilon_{\text{Al}}}{N_{\text{Mg}} + N_{\text{Al}}},
  \label{eq:hof}
\end{equation}
where $E_{\text{tot}}$ is the total energy of the system,
$N_{\text{Mg}}$ and $N_{\text{Al}}$ are the numbers of Mg and Al atoms
in the system, $\varepsilon_{\text{Mg}}$ and $\varepsilon_{\text{Al}}$
are the total energies per atom for Mg and Al in their ideal bulk
structures, respectively.  Table~\ref{tab:Mg-Al-in-B1} lists some of
the material properties the new MEAM potential reproduces for a MgAl
compound in a B1 structure compared with the predictions of the DFT
calculations.  Due to our emphasis on the first three properties
during the construction process, the last two columns show some
discrepancies between the MEAM and DFT results.
\begin{table}[!tbp]
  \caption{\label{tab:Mg-Al-in-B1} Elastic parameters for
    MgAl in a B1 structure from the MEAM and DFT calculations. The units of
    the heat of formation per atom $H_{\text{f}}$ and the equilibrium
    atomic volume $V_0$ are eV and $\AA^3$, respectively. The units
    of the bulk modulus $B_0$ and all elastic constants $C_{ij}$ are GPa.}
  \begin{ruledtabular}
    \begin{tabular}{rccccc}
      Method & $H_{\text{f}}$ & $V_0$ & $B_0$ & $C_{44}$ & $(C_{11}-C_{12})/2$\\
      \hline
      DFT  & 0.4575 & 22.4 &   38.4 &  -39.1 &   40.6\\
      MEAM & 0.4575 & 22.4 &   38.4 &  -14.3 &   29.8
    \end{tabular}
  \end{ruledtabular}
\end{table}

\begin{figure}[!tbp]
  \includegraphics[width=1\columnwidth]{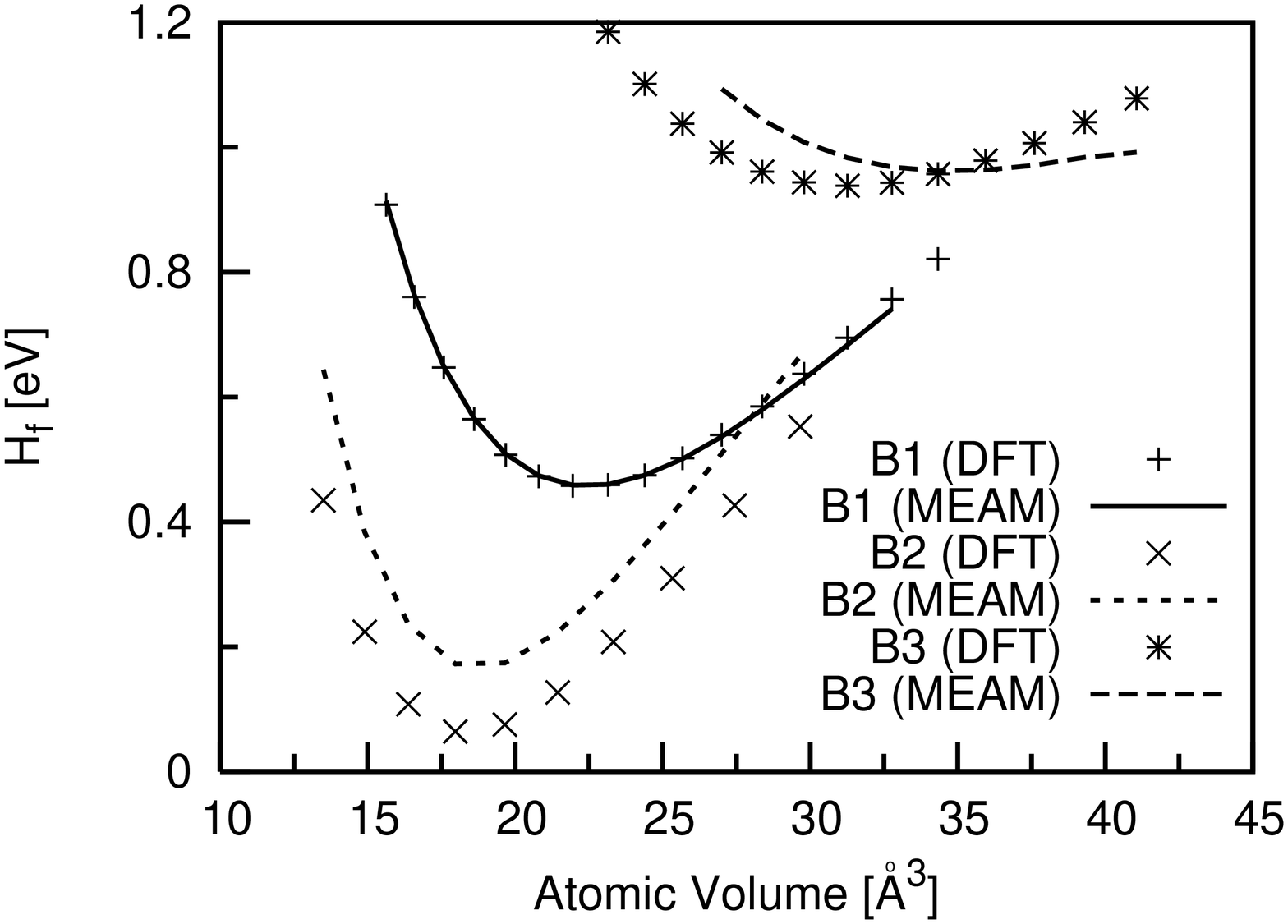}
  \caption{\label{fig:AlMg_struct} The heat of formation per atom for
    MgAl alloys in the B1, B2 and B3 crystal structures.}
\end{figure}

Fig.~\ref{fig:AlMg_struct} shows the heat of formation per atom
$H_{\text{f}}$ for the B1, B2 and B3 structures compared with the
results from the DFT calculations.  The B1 (cubic rock salt) structure
was used as the reference structure for the Mg-Al alloy MEAM potential.
Fig.~\ref{fig:AlMg_struct} shows that the reference structure is not
the most stable structure in Mg-Al binary systems. Again, the sequence
of the structures in the order of stability is predicted correctly by
the new MEAM potential for the Mg-Al alloy system. The relative
cohesive energies, with respect to the one for the reference
structure, are also in good agreement with the DFT calculations.  The
equilibrium atomic volume and bulk modulus of Mg-Al in the B1
structure are reproduced almost exactly.  Note that the abscissa of
the plot in Fig.~\ref{fig:AlMg_struct} is the actual volume instead of
the volume ratio used in Fig.~\ref{fig:Al_struct} and
Fig.~\ref{fig:Mg_struct}.

\begin{figure}[!tbp]
  \includegraphics[width=1\columnwidth]{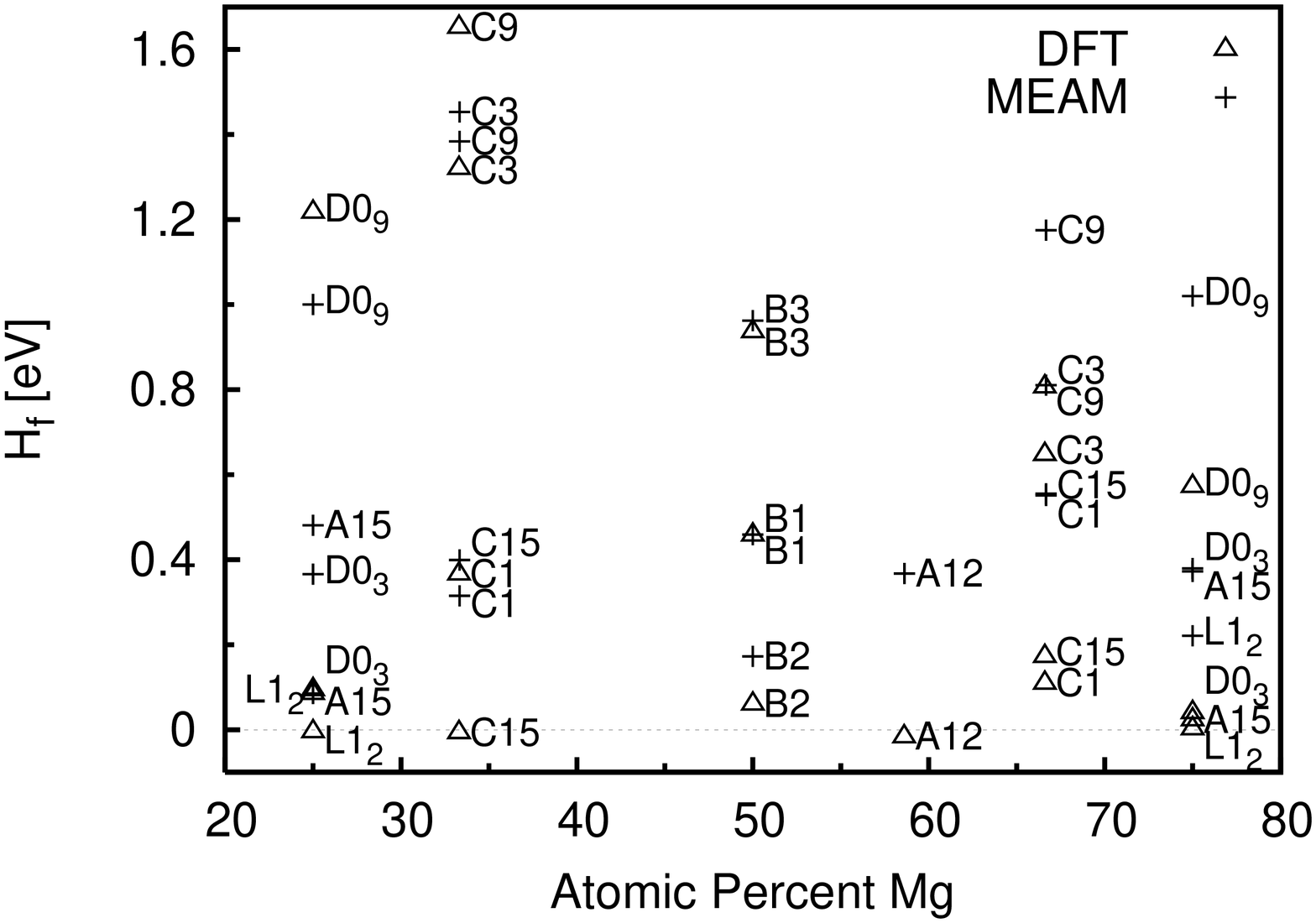}
  \caption{\label{fig:AlMg-alloy-Hof} The heat of formation per atom
    for Mg-Al alloys in various intermetallic phases with different
    stoichiometric coefficients. The results obtained from the new
    Mg-Al MEAM potentials are compared with the DFT calculations. The
    structure names are written next to the symbols (open triangles
    for DFT and crosses for MEAM).}
\end{figure}

To further demonstrate the validity of our new potentials, we also
computed the heat of formation per atom for many intermetallic phases
of Mg-Al alloys.  The total energy values in Eq.~\ref{eq:hof} of B1,
B2, B3, C1, C3, C9, C15, D0$_3$, D0$_9$, A15, L1$_2$ and A12
structures were evaluated at the optimal atomic volume for each
structure.  The results from the MEAM calculations, compared with the
ones from the DFT calculations, are summarized in
Fig.~\ref{fig:AlMg-alloy-Hof}.  Although the Mg and Al atoms in these
intermetallic phases are in a chemical environment very different from
the one in the reference structure (B1), the agreement between MEAM
and DFT is quite satisfactory. In most cases, MEAM preserves the order
of stability predicted by DFT.  The differences in the heat of
formation per atom from MEAM and DFT are less than 0.5~eV at most.
However, we note that the MEAM failed to predict that the formation of
one of the experimentally observed Mg-Al alloy
structures\cite{Singh03:Mg-Al-exp, okamoto98:_al_mg},
$\gamma$(Mg$_{17}$Al$_{12}$) denoted as the $A12$ structure in
Fig.~\ref{fig:AlMg-alloy-Hof}, as an exothermic process.  In
comparison, our DFT calculation correctly predicted the $H_{\text{f}}$
for this structure to be a negative value (-0.017~eV).

\subsection{Surfaces}

\subsubsection{Surface formation energies}

Semi-infinite surface is one of the simplest forms of defects. To test
the transferability of the new MEAM potentials, surface formation
energies for several different surfaces are computed.
Surface formation energy per unit surface area $E_{\text{surf}}$ is
defined as
\begin{equation}
  E_{\text{surf}} = \left(E_{\text{tot}} - N \varepsilon \right) / A,
\end{equation}
where $E_{\text{tot}}$ is the total energy of the structure with a
surface, $N$ is the number of atoms in the structure, $\varepsilon$ is
the total energy per atom in the bulk, and $A$ is the surface area.
Table~\ref{tab:E_surf} shows the surface formation energies of many
different surfaces constructed from fcc Al and hcp Mg crystals.
Results from the present MEAM potentials are in good agreement with
the DFT calculations, representing a significant improvement over two
of the previously published MEAM potentials
\cite{lee03:_semiem_cu_ag_au_ni, hu01:_analy_hcp}. However, our MEAM
potentials and the EAM potentials by \citet{Liu:SURF-v373-1997}
exhibit comparable levels of validity: our new MEAM potentials perform
better for Mg surfaces while Liu's EAM potentials give better
agreement for Al surfaces.
\begin{table}[!tbp]
  \caption{\label{tab:E_surf} Surface formation energies for fcc Al
    and hcp Mg. The units are mJ/$\text{m}^2$. The second column
    indicates if the structure was relaxed.
    Comparisons with other previously developed MEAM potentials
    are also given.}
  \begin{ruledtabular}
    \begin{tabular}{ccccccc}
      \multirow{2}{*}{Surface} & \multirow{2}{*}{Relaxed} &
      \multirow{2}{*}{MEAM\footnotemark[1]} & \multirow{2}{*}{DFT} &
      \multicolumn{3}{c}{Others} \\
      \cline{5-7}
      & & & & Ref.~\onlinecite{lee03:_semiem_cu_ag_au_ni} 
      & Ref.~\onlinecite{hu01:_analy_hcp} 
      & Ref.~\onlinecite{Liu:SURF-v373-1997}\footnotemark[2] \\
      \hline
      Al(111) & No & 737 & 992 & & & 913 \\
      Al(111) & Yes & 731 & 988 & 629 & & 912 \\
      Al(110) & No & 1068 & 1371 & & & 1113 \\
      Al(110) & Yes & 1035 & 1349 & 948 & & 1107 \\
      Al(100) & No & 1025 & 1213 & & & 1012 \\
      Al(100) & Yes & 1025 & 1212 & 848 & & 1002 \\
      \hline
      Mg(0001) & No & 604 & 638 & & & 500 \\
      Mg(0001) & Yes & 595 & 637 & & 310 & 499 \\
      Mg(10$\bar{1}$0) & No & 642 & 855 & & & 517 \\
      Mg(10$\bar{1}$0) & Yes & 523 & 846 & & 316 & 515 \\
    \end{tabular}
  \end{ruledtabular}
  \footnotetext[1]{The present MEAM potential.}
  \footnotetext[2]{Calculated using EAM parameters extracted from
    Ref.~\onlinecite{Liu:SURF-v373-1997}.}
\end{table}

\subsubsection{Stacking fault energies}

Stacking fault is another kind of structure that occurs frequently in
real materials and provides a good test environment for newly
developed semiempirical potentials.  Stacking fault energy per unit
area is defined by
\begin{equation}
  E_{\text{sf}} = \left(E_{\text{tot}} - N \varepsilon \right) / A,
\end{equation}
where $E_{\text{tot}}$ is the total energy of the structure with a
stacking fault, $N$ is the number of atoms in the system,
$\varepsilon$ is the total energy per atom in the bulk, and $A$ is the
unit cell area that is perpendicular to the stacking fault.  

For Al, three stacking fault types from \citet{hirth82:_disloc} were
examined and the results are listed in Table~\ref{tab:sf_Al}. For the
case of stacking fault type $I$, our MEAM result is in a good
agreement with the available experimental value, even though our DFT
result is lower than the experimental value.  In all cases considered,
the present MEAM potential shows better overall agreements with DFT
calculations compared with the EAM potential by
\citet{Liu:SURF-v373-1997}.
\begin{table}[!tbp]
  \caption{\label{tab:sf_Al} Stacking fault energies for Al.
    Results from the present MEAM and DFT calculations are compared. 
    Stacking fault energies per unit area are given in mJ/$\text{m}^2$.}
  \begin{ruledtabular}
    \begin{tabular}{cccrrr}
      \multirow{2}{*}{Element} & \multirow{2}{*}{Fault} &
      \multirow{2}{*}{Relaxed} &
      \multicolumn{2}{c}{MEAM (DFT)} &
      \multirow{2}{*}{Exp.\footnotemark[2]} \\
      \cline{4-5}
      & & & Present & Other\footnotemark[1] & \\
      \hline
      Al & $I$ & No & 150 (136) & 169 & \\
      Al & $I$ & Yes & 146 (133) & 142 & 140-160 \\
      Al & $E$ & No & 150 (135) & 169 & \\
      Al & $E$ & Yes & 148 (133) & 154 & \\
      Al & $T$ & No & 75 (62) & 84 & \\
      Al & $T$ & Yes & 74 (61) & 77 & \\
    \end{tabular}
  \end{ruledtabular}
  \footnotetext[1]{Calculated using EAM parameters extracted from
    Ref.~\onlinecite{Liu:SURF-v373-1997}.}
  \footnotetext[2]{Experimental results from
    Ref.~\onlinecite{hirth82:_disloc}.}
  \footnotetext{$I=ABCBCABC$}
  \footnotetext{$E=ABCABCBABCABC$}
  \footnotetext{$T=ABCABCABACBACB$}
\end{table}

For Mg, four stacking fault types from the calculation of
\citet{chetty97:_stack} were examined. Total energy calculations for
$I_1$, $I_2$, $T_2$, and $E$ stacking fault types were performed using
both DFT and MEAM calculations. The results are compared in
Table~\ref{tab:sf_Mg}.  The present MEAM potential shows a substantial
improvement over the previously published MEAM potential by
\citet{hu01:_analy_hcp}. The stacking fault energies are consistently
underestimated by the present MEAM potentials compared to the results
of the DFT calculations, while the results by the EAM potential from
Ref.~\onlinecite{Liu:SURF-v373-1997} are consistently overestimated.
\begin{table}[!tbp]
  \caption{\label{tab:sf_Mg} Stacking fault energies for Mg.
    Results from the present MEAM and DFT calculations are compared. 
    Stacking fault energies per unit area are given in
    mJ/$\text{m}^2$. Comparisons with other previously developed MEAM potentials
    are also given.}
  \begin{ruledtabular}
    \begin{tabular}{ccccccc}
      Element & Fault & MEAM\footnotemark[1] & EAM\footnotemark[2] &
      AMEAM\footnotemark[3] & DFT\footnotemark[4] & DFT\footnotemark[5] \\
      \hline
      Mg & $I_1$ & 7 & 27 & 4 & (18) & (20.9) \\
      Mg & $I_2$ & 15 & 54 & 8 & (37) & (43.7) \\
      Mg & $T_2$ & 15 & 54 & -- & (45) & (51.3) \\
      Mg & $E$   & 22 & 81 & 12 & (61) & (68.1) \\
    \end{tabular}
  \end{ruledtabular}
  \footnotetext[1]{The present MEAM potential.}
  \footnotetext[2]{Calculated using EAM parameters extracted from
    Ref.~\onlinecite{Liu:SURF-v373-1997}.}
  \footnotetext[3]{AMEAM results in Ref.~\onlinecite{hu01:_analy_hcp}}
  \footnotetext[1]{DFT results from the present study.}
  \footnotetext[5]{DFT results in Ref.~\onlinecite{chetty97:_stack}} 
  \footnotetext{$I_1=ABABAB CBCBCB$}
  \footnotetext{$I_2=ABABAB CACACB$}
  \footnotetext{$T_2=ABABAB CBABAB$}
  \footnotetext{$E=ABABAB CABABAB$}
\end{table}

\subsubsection{Adsorption on surfaces}

The adsorption energy of a single adatom $E_{\text{ads}}$ is given by
\begin{equation}
  E_{\text{ads}} = E_{\text{tot}} - E_{\text{surf}} - E_{\text{atom}},
\end{equation}
where $E_{\text{tot}}$ is the total energy of the structure with the
adatom adsorbed on the surface, $E_{\text{surf}}$ is the total energy
of the surface without the adatom, and $E_\text{atom}$ is the total
energy of an isolated atom. We placed single Al and Mg atoms at the
fcc and hcp sites (see, e.g., \citet{Lovvik98:H-on-Pd(111)} for the
definition of these sites) on Al(111) and Mg(0001) surfaces.  The
entire structures were then relaxed to determine the adsorption
energies.  The height of the relaxed adatom is measured from the
farthest atom (the one least affected by the adsorption) in the top
surface layer.  Table~\ref{tab:adsorb} shows the results obtained from
the MEAM and DFT calculations. Even though there are small
quantitative discrepancies, the qualitative agreement between these
two sets of data is quite satisfactory. For instance, DFT calculations
predict that on Al(111) surfaces the adsorption of the Al atoms (same
kind) is stronger (bigger adsorption energies and shorter heights)
than the adsorption of Mg atoms (different kind). On the other hand,
DFT calculations predict that on Mg(0001) surfaces, the adsorption of
the Mg atoms (same kind) is weaker (smaller adsorption energies and
longer heights) than the adsorption of Al atoms (different kind). Both
of these features are clearly demonstrated by the new MEAM potentials.
\begin{table}[!tbp]
  \caption{\label{tab:adsorb} Adsorption energies $E_{\text{ads}}$ and
    optimized height above the surface from MEAM calculations. Results 
    from DFT calculations are given in parentheses. Units are eV and
    \AA\ for energies and heights, respectively.}
  \begin{ruledtabular}
    \begin{tabular}{cccc}
      Surface & Adatom (site) & $E_{\text{ads}}$ & Height\\
      \hline
      Al(111)  & Al (hcp) & -2.64 (-3.29) & 2.19 (2.05) \\
      Al(111)  & Al (fcc) & -2.67 (-3.26) & 2.17 (2.08) \\
      Al(111)  & Mg (hcp) & -1.70 (-1.09) & 2.46 (2.35) \\
      Al(111)  & Mg (fcc) & -1.70 (-1.07) & 2.47 (2.35) \\
      \hline
      Mg(0001) & Al (hcp) & -2.17 (-2.68) & 2.11 (2.16) \\
      Mg(0001) & Al (fcc) & -2.17 (-2.68) & 2.09 (2.14) \\
      Mg(0001) & Mg (hcp) & -1.43 (-0.81) & 2.17 (2.28) \\
      Mg(0001) & Mg (fcc) & -1.49 (-0.82) & 2.37 (2.27) \\
    \end{tabular}
  \end{ruledtabular}
\end{table}

\subsection{Point defects}

\subsubsection{Vacancy}

The formation energy of a single vacancy $E_{\text{f}}^{\text{vac}}$
is defined as the energy cost to create a vacancy:
\begin{equation}
  E_{\text{f}}^{\text{vac}} = E_{\text{tot}}[N] - N\varepsilon,
  \label{eq:E^f_v}
\end{equation}
where $E_\text{tot}[N]$ is the total energy of a system with $N$ atoms
containing a vacancy and $\varepsilon$ is the energy per atom in the
bulk.  Table~\ref{tab:vac} shows the formation energy of single
vacancies for fcc Al and hcp Mg obtained from the MEAM and DFT
calculations.  The new MEAM potentials reproduced a DFT value for Al
vacancy formation energy very well, although the value for Mg was
estimated somewhat low.  Furthermore, the present MEAM potentials
reproduce the correct amount of reduction in volume due to the
formation of a vacancy.  This also represents a substantial
improvement over the existing MEAM potentials.
\begin{table}[!tbp]
  \caption{\label{tab:vac} Calculated single vacancy properties. 
    Single vacancy formation energy $E_{\text{f}}^{\text{vac}}$ and
    formation volume $\Omega_v$ values are obtained from the relaxed
    structures containing single vacancies. Here $\Omega_0$ is the
    bulk atomic volume. All energy values are 
    listed in eV.  The results from the MEAM calculations
    are compared with the results from the DFT calculations given inside 
    the parentheses.}
  \begin{ruledtabular}
    \begin{tabular}{cccccc}
      \multirow{2}{*}{Element} & \multicolumn{2}{c}{$E_{\text{f}}^{\text{vac}}$} 
      & & \multicolumn{2}{c}{$\Omega_v/\Omega_0$} \\
      \cline{2-3} \cline{5-6}
      & Present & Others & & Present & Others \\
      \hline
      Al & 0.68 (0.67) & 0.68\footnotemark[1], 0.68\footnotemark[2]  
      & & 0.66 (0.76) & 0.72\footnotemark[1], 0.61\footnotemark[2] \\
      \hline
      Mg & 0.58 (0.82) & 0.59\footnotemark[3], 0.87\footnotemark[2] 
      & & 0.76 (0.75) & 0.83\footnotemark[3], 0.88\footnotemark[2] \\
    \end{tabular}
  \end{ruledtabular}
  \footnotetext[1]{MEAM results in Ref.~\onlinecite{lee03:_semiem_cu_ag_au_ni}}
  \footnotetext[2]{Calculated using EAM parameters extracted from
    Ref.~\onlinecite{Liu:SURF-v373-1997}.}
  \footnotetext[3]{AMEAM results in Ref.~\onlinecite{hu01:_analy_hcp}}
\end{table}

\subsubsection{Interstitial point defects}

\begin{table}[!tbp]
  \caption{\label{tab:interstitials} The formation energies of various
    kinds of interstitial point defects in Al and Mg. All energy
    values are given in eV.  The results from the MEAM calculations
    are compared with the results from the DFT calculations given inside 
    the parentheses.}
  \begin{ruledtabular}
    \begin{tabular}{ccc}
      Bulk (structure) & Interstitial (site) & MEAM (DFT)\\ 
      \hline
      Al (fcc) & Al (dumbbell) & 2.32 (2.94)\\
      % Al(fcc) & Al (Dumbbell) & Full & 2.49\footnotemark[1](1.579)\footnotemark[2]\\
      Al (fcc) & Al (octahedral)  & 2.91 (3.06)\\
      % Al (fcc) & Al (octahedral) & Full & (1.79)\footnotemark[2]\\
      Al (fcc) & Al (tetrahedral) & 3.14 (3.68)\\ 
      % Al (fcc) & Al (tetrahedral) & Full & (1.978\footnotemark[3])\\ 
      Al (fcc) & Mg (octahedral)  & 2.77 (3.79)\\ 
      Al (fcc) & Mg (tetrahedral) & 5.09 (4.25)\\
      \hline
      Mg (hcp) & Mg (octahedral)  & 1.29 (2.36)\\
      Mg (hcp) & Mg (tetrahedral) & 1.53 (2.35)\\
      Mg (hcp) & Al (octahedral)  & 2.13 (1.97)\\
      Mg (hcp) & Al (tetrahedral) & 2.79 (2.11)\\
    \end{tabular}
  \end{ruledtabular}
  % \footnote[1]{\citet{lee03:_semiem_cu_ag_au_ni}}
  % \footnote[2]{\citet{PhysRevB.55.4941}}
\end{table}
The formation energy of an interstitial point defect
$E_{\text{f}}^{\text{int}}$ is given by
\begin{equation}
  E_{\text{f}}^{\text{int}} = E_{\text{tot}}[N+A] - E_{\text{tot}}[N] 
  - \varepsilon_{\text{A}}
  \label{eq:Ef_int}
\end{equation}
where $E_\text{tot}[N]$ is the total energy of a system with $N$ (Mg
or Al) atoms, $E_\text{tot}[N+A]$ is the total energy of a system with
$N$ atoms plus one atom of type-$A$ (Mg or Al) inserted at one of the
interstitial sites, and $\varepsilon_{\text{A}}$ is the total energy
per atom of type-$A$ in its most stable bulk structure. Note that the
inserted atom $A$ can be the same type as the matrix, in which case
the point defect becomes a so-called self-interstitial defect.
Interstitial atom formation energies were calculated for Al and Mg at
octahedral, tetrahedral, and dumbbell sites.  Atomic position and
volume relaxation were performed.  The results of these calculations
are listed in Table~\ref{tab:interstitials}, to be compared with the
results from the DFT calculations. DFT results are well reproduced in
general.  According to the present calculations, the most stable form
of a self-interstitial defect for fcc Al crystal is a dumbbell along
the [100] direction, in agreement with the DFT results and an
experimental observation by \citet{PhysRevB.55.4941}.  The new MEAM
potentials, however, failed to reproduce the results of the DFT
calculations for the self-interstitial defects in a hcp Mg crystal.
Our new MEAM potential indicates that the octahedral site will be more
stable than the tetrahedral site, while the DFT calculations predict that
both sites will have nearly the same formation energies.  In both of
the heterogeneous interstitial defects, the new MEAM potentials
produce the same relative stability of different interstitial sites
with the DFT calculations.  However, Table~\ref{tab:interstitials} should
not be used to predict the most stable interstitial defects; it was
not the purpose of the present work to perform an exhaustive search to
draw such conclusions.

\subsubsection{Substitutional point defects}

The formation energy of a substitutional point defect
$E_{\text{f}}^{\text{sub}}$, in the case of the substitution of an Al
atom with a Mg atom, is defined by
\begin{equation}
  E_{\text{f}}^{\text{sub}} = E_{\text{tot}}[\text{Mg}_{\text{Al}}] 
  - E_{\text{tot}}[\text{Al}_{\text{Al}}] 
  - \varepsilon_{\text{Mg}} + \varepsilon_{\text{Al}}
  \label{eq:E^f_sub}
\end{equation}
where $E_{\text{tot}}[\text{Mg}_{\text{Al}}]$ is the total energy of a
system of Al atoms plus one Mg atom that replaced an Al atom,
$E_{\text{tot}}[\text{Al}_{\text{Al}}]$ is the total energy of the
original system of Al atoms without a defect,
$\varepsilon_{\text{Mg}}$ and $\varepsilon_{\text{Al}}$ are the total
energies per atom for Mg and Al in their ideal bulk structures.  The
formation energy of a substitutional point defect for other cases can
be defined similarly.  Table~\ref{tab:E^f_sub} shows the results of
substitutional defect calculations using the MEAM potentials and the
DFT method.  The new MEAM potentials predict correctly that
substituting a Mg atom in a hcp structure with an Al atom costs more
energy than the reverse as indicated by the DFT results, although the
formation energies in both cases are larger than the values from the
DFT calculations. 
\begin{table}[!tbp]
  \caption{\label{tab:E^f_sub} The formation energies of substitutional 
    point defects in Al and Mg. All energy values are given in eV.}
  \begin{ruledtabular}
    \begin{tabular}{cccc}
      Bulk (structure) & Substitute & MEAM (DFT) \\
      \hline
      Al (fcc) & Mg & 0.38 (0.06) \\
      Mg (hcp) & Al & 0.55 (0.09) \\
    \end{tabular}
  \end{ruledtabular}
\end{table}

\subsection{Molecular dynamics simulations}

To validate the new potentials for molecular dynamics simulations, we
calculated the melting temperatures of pure Al and Mg crystals. We
followed the procedure prescribed by
\citet{Morris:1994:PhysRevB.49.3109} to establish co-existence of
solid and liquid phases to determine the melting temperatures. We
obtained 930 K for pure Al crystal, which is in excellent agreement
with the experimental value of 933 K.  For Mg crystal, however, the
two-phase method did not give a satisfactory result: 530 K compared to
the experimental value of 923 K. To compare with other potentials, we
followed a single-phase method as described by \citet{Kim:1994:PRL},
in which the temperature is increased at a constant rate and the
specific heat of the system is monitored. Using this method, we
obtained 780 K as the melting temperature of Mg crystals. This result
is comparable to 745 K obtained by \citet{Liu:1996:MSMSE} using an EAM
potential and following a similar method to compute the melting
temperature.

The difficulty in computing accurate melting temperatures for hcp
metals, such as Mg, using semiempirical potentials is well
documented\cite{Sun:2006:PRB}.  We believe that this is related to a
small energy difference between hcp and fcc structures in our MEAM
potential for Mg (see Fig.~\ref{fig:Mg_struct}) and instability of hcp
structures in MEAM\cite{Mae:2002:MSMSE}.  Further detailed
investigation of this subject is beyond the scope of the present work
and will be reported in separate papers.

\section{Conclusions}
\label{sec:Conclusion}

In this study we developed a new set of MEAM potentials for Al, Mg and
their alloy systems using first-principles calculations based on DFT.
The validity and transferability of the new MEAM potentials were
tested rigorously by calculating physical properties of the Mg-Al
alloy systems in many different atomic arrangements such as bulk,
surface, and point defect structures.  The new MEAM potentials show
a significant improvement over the previously published potentials,
especially for the surface formation, stacking faults, and point
defect calculations. The new Mg-Al alloy potentials, however, failed
to predict the stability of the $\gamma$(Mg$_{17}$Al$_{12}$) alloy
intermetallic phase, suggesting the need for further improvements.

\section{Acknowledgment}

The authors are grateful to the Center for Advanced Vehicular Systems
at Mississippi State University for supporting this study.  Computer
time allocation has been provided by the High Performance Computing
Collaboratory (HPC$^2$) at Mississippi State University.  This work
has also been supported in part by the US Department of Defense under
the CHSSI MBD-04 (Molecular Packing Software for \textit{ab initio}
Crystal Structure and Density Predictions) project (S.G.K.) and by the U.S.
Department of Energy, Office of Basic Energy Sciences (M.I.B.).

\bibliography{MgAl,DFT}

\end{document}